\documentclass[
    ,final            
  ]
  {aipproc}
  
\layoutstyle{6x9}

\usepackage{graphicx}
\usepackage{bm}
\usepackage{subfigure}

\begin{document}

\title{Anisotropic hydrodynamics \\ and the early-thermalization puzzle}


\author{Wojciech Florkowski}{
  address={Institute of Physics, Jan Kochanowski University, PL-25406~Kielce, Poland, \\
and The H. Niewodnicza\'nski Institute of Nuclear Physics, Polish Academy of Sciences, PL-31342 Krak\'ow, Poland}
}

\author{Radoslaw Ryblewski}{
  address={The H. Niewodnicza\'nski Institute of Nuclear Physics, Polish Academy of Sciences, PL-31342 Krak\'ow, Poland}
}

\begin{abstract}
The framework of anisotropic hydrodynamics is used in 3+1 dimensions to analyze behavior of matter produced in ultra-relativistic heavy-ion collisions. The model predictions for the hadronic transverse-momentum spectra, directed and elliptic flows, and the HBT radii are presented. We show that the effects of the initial anisotropy of pressure may be compensated by appropriate adjustment of the initial energy density.  In this way, the final hadronic observables become insensitive to the early stage dynamics and the early thermalization puzzle may be circumvented.
\end{abstract}

\maketitle


\section{Introduction}

Soft-hadronic observables measured in the ultra-relativistic heavy-ion experiments are well reproduced by perfect-fluid hydrodynamics \cite{Florkowski:2010zz} or by viscous  hydrodynamics with a small viscosity to entropy ratio \cite{Chaudhuri:2006jd,Dusling:2007gi,Luzum:2008cw,Song:2007fn,Bozek:2009dw,Schenke:2010rr}. Nevertheless, the use of such approaches at the very early stages of the collisions encounters important conceptual difficulties. Thermalization times shorter than a fraction of a fermi (used quite often in the perfect-fluid approaches) cannot be explained within the microscopic models of the collisions. On the other hand, viscous hydrodynamics is based on an implicit assumption that one can make an expansion around the isotropic background. If the shear correction is large, a new framework is needed. This has triggered developments of the reorganization of viscous hydrodynamics
in which one incorporates the possibility of large momentum-space anisotropies into the
leading order of the approximation. Such a new framework has been introduced in \cite{Florkowski:2010cf,Martinez:2010sc,Ryblewski:2010bs,Martinez:2010sd,Ryblewski:2011aq,Martinez:2012tu,Ryblewski:2012rr} and we refer to it below as to the {\it anisotropic hydrodynamics}.

\section{Anisotropic hydrodynamics}

Anisotropic hydrodynamics is based on the equations
\begin{eqnarray}
\partial_\mu T^{\mu \nu} &=& 0, \label{enmomcon} \\
\partial_\mu \sigma^{\mu} &=& \Sigma, \label{engrow}
\end{eqnarray}
which express the energy-momentum conservation and entropy production laws. The energy-momentum tensor $T^{\mu \nu}$ has the structure
\begin{eqnarray}
T^{\mu \nu} = \left( \varepsilon  + P_{\perp}\right) U^{\mu}U^{\nu} - P_{\perp} \, g^{\mu\nu} - (P_{\perp} - P_{\parallel}) V^{\mu}V^{\nu}, 
\label{Taniso}
\end{eqnarray}
where $P_{\parallel}$ is the longitudinal pressure and $P_{\perp}$ is the transverse pressure. In the limit $P_{\parallel}=P_{\perp}=P$, Eq.~(\ref{Taniso}) reproduces the energy-momentum tensor of the perfect fluid. Similarly, the entropy production law (\ref{engrow}) is reduced to the entropy conservation law, if we take $\Sigma=0$. The four-vector $U^{\mu}$ is the four-velocity of the fluid, while $V^{\mu}$ is the four-vector whose spatial part is parallel to the beam ($z$) axis. In the general case, we use the parameterizations $U^\mu = (u_0 \cosh \vartheta, u_x, u_y, u_0 \sinh \vartheta)$ and $V^\mu = (	 \sinh \vartheta, 0, 0,  \cosh \vartheta)$, where $u_x$ and $u_y$ are the transverse components of the four-velocity field and $\vartheta$ is the longitudinal fluid rapidity. The entropy flux $\sigma^{\mu}$ is defined by the formula
\begin{equation}
\sigma^{\mu} = \sigma \, U^\mu,
\label{Saniso}
\end{equation}
where $\sigma$ is the non-equilibrium entropy density.

One can show \cite{Florkowski:2010cf} that instead of $P_{\parallel}$ and $P_{\perp}$ it is more convenient to use the entropy density $\sigma$ and the anisotropy parameter $x$ as two independent variables (one may use the approximation $P_{\parallel}/P_{\perp} = x^{-3/4}$). Similarly to standard hydrodynamics with vanishing baryon chemical potential, the energy density $\varepsilon$, the entropy density  $\sigma$, and the anisotropy parameter $x$ are related through the \textit{generalized} equation of state $\varepsilon=\varepsilon(\sigma, x)$. In our model we use the following ansatz \cite{Ryblewski:2010bs}
\begin{eqnarray}
\varepsilon (x,\sigma)&=&  \varepsilon_{\rm qgp}(\sigma) r(x), \label{epsilon2b}  \\ \nonumber 
P_\perp (x,\sigma)&=&  P_{\rm qgp}(\sigma) \left[r(x) + 3 x r^\prime(x) \right], \label{PT2b}   \\ \nonumber 
P_\parallel (x,\sigma)&=&  P_{\rm qgp}(\sigma) \left[r(x) - 6 x r^\prime(x) \right]. \label{PL2b} 
\end{eqnarray}
where the functions $\varepsilon_{\rm qgp}$ and  $P_{\rm qgp}$  define the realistic QCD equation of state constructed in Ref. \cite{Chojnacki:2007jc}. The function $r(x)$ characterizes  properties of the fluid which exhibits the pressure anisotropy $x$.  Here we use the form introduced in \cite{Florkowski:2010cf}
\begin{equation}
r(x) = \frac{ x^{-\frac{1}{3}}}{2} \left[ 1 + \frac{x \arctan\sqrt{x-1}}{\sqrt{x-1}}\right].
\label{RB}
\end{equation}
In the isotropic case $x = 1$, $r(1)=1$, $r^\prime(1)=0$, and Eq.~(\ref{epsilon2b}) is reduced to the standard equation of state used in \cite{Chojnacki:2007jc}.

The function $\Sigma$ in Eq.~(\ref{engrow}) defines the entropy source. We use the form proposed in \cite{Florkowski:2010cf}
\begin{equation}
\Sigma(\sigma,x) = \frac{(1-\sqrt{x})^{2}}{\sqrt{x}}\frac{\sigma}{\tau_{\rm eq}},
\label{entropys}
\end{equation}
where the time-scale parameter $\tau_{\rm eq}$ controls the rate of equilibration\footnote{In this work we use the constant value $\tau_{\rm eq}$ = 1 fm. In Refs. \cite{Martinez:2010sc,Martinez:2010sd,Martinez:2012tu} the medium dependent $\tau_{\rm eq}$ was used, which was inversely proportional to the typical transverse momentum scale in the system. If the constant value of $\tau_{\rm eq}$ is used, the system approaches the perfect fluid behavior for $\tau \gg \tau_{\rm eq}$.}. In the limit of small anisotropy Eq.~(\ref{entropys}) is consistent with the quadratic form of the entropy production in the Israel-Stewart theory. Far from equilibrium, hints for the form of $\Sigma$ are lacking, although we may expect some suggestions from the AdS/CFT correspondence \cite{Heller:2011ju}. Thus, for large anisotropies the formula (\ref{entropys}) should be treated as an assumption defining the dynamics of the system. 

\begin{figure}[t]
\includegraphics[angle=0,width=0.5\textwidth]{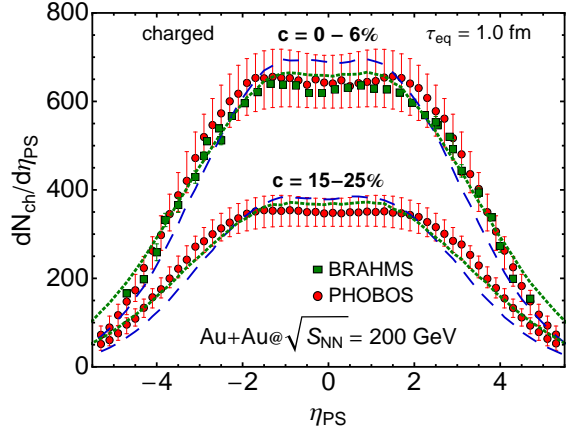}
\caption{\small (Color online) The pseudorapidity distribution of charged particles for \mbox{$\tau_{\rm eq}=1.0$ fm}. The results are obtained for $x_{\rm 0}=100$ (dashed blue lines) and $x_{\rm 0}=0.032$ (dotted green lines), and for the two centrality classes: $c=0-6$\% ($b=2.48$ fm) and $c=15-25$\% ($b=6.4$ fm). The model results are compared to experimental data from PHOBOS \cite{Back:2002wb} (red dots) and BRAHMS \cite{Bearden:2001qq} (green squares).}
\label{fig:etadistr_RHIC}
\end{figure}

\begin{figure}[t]
\includegraphics[angle=0,width=0.5\textwidth]{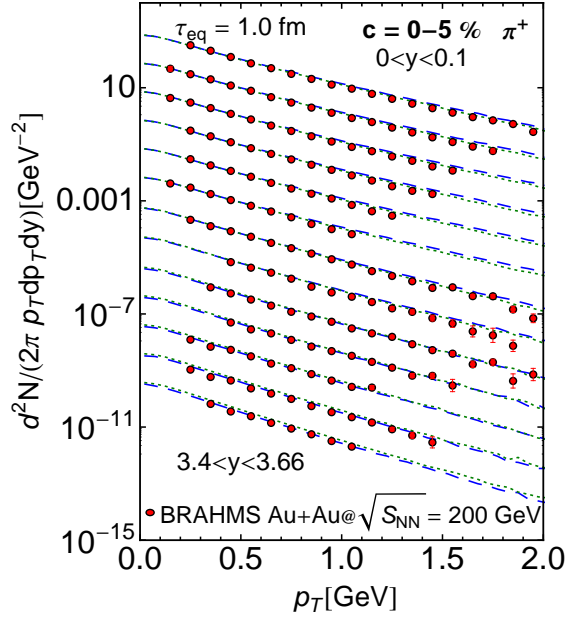}
\caption{\small (Color online) Transverse momentum spectra of $\pi^{+}$'s for the centrality class $c=0-5$\% ($b=2.26$ fm) and for different rapidity windows; $x_{\rm 0}=100$ (dashed blue lines) and $x_{\rm 0}=0.032$ (dotted green lines), and $\tau_{\rm eq}= 1.0$ fm. The model results are compared to the experimental data from BRAHMS \cite{Bearden:2004yx}. The spectra for different rapidity windows are successively rescaled down by factor $0.1$. 
}
\label{fig:rapptdistr_RHIC3}
\end{figure}

\section{Initial conditions and freeze-out}
\label{sec_ini}
In the general 3+1 case we have to solve Eqs. (\ref{enmomcon}) and (\ref{engrow}) for $\sigma$, $x$, $u_x$, $u_y$, and $\vartheta$, which depend on $\tau,{\bf x}_\perp=(x,y)$, and $\eta$ ($\tau$ is the proper time and $\eta$ is the space-time rapidity). We fix the initial starting time to $\tau_{\rm 0} =0.25$ fm. Similarly to other hydrodynamic calculations, we assume that there is no initial transverse flow, $u_x(\tau_{\rm 0},{\bf x}_\perp,\eta) = u_y(\tau_0,{\bf x}_\perp,\eta) = 0$ and that the initial longitudinal rapidity of the fluid is equal to space-time rapidity, $\vartheta(\tau_0,{\bf x}_\perp,\eta) = \eta$. In this text we present the results for two scenarios: i) the initial source is strongly {\it oblate} in the momentum space, $x(\tau_0,{\bf x}_\perp,\eta) =100$, and ii) the source is {\it prolate} in momentum space, $x(\tau_0,{\bf x}_\perp,\eta) = 0.032$; the latter value is chosen because $r(100) = r(0.032)$. 

The initial entropy density profile has the form
\begin{equation}
 \sigma_0(\eta,{\bf x}_\perp) = \sigma(\tau_0,\eta,{\bf x}_\perp) = \varepsilon_{\rm gqp}^{-1} 
\left[ \varepsilon_{\rm i} \, \tilde{\rho}(b,\eta,{\bf x}_\perp) \right],
\label{sig2}
\end{equation}
where $b$ is the impact parameter, and $\tilde{\rho}(b,\eta,{\bf x}_\perp)$ is the normalized density of sources, $\tilde{\rho}(b,\eta,{\bf x}_\perp) = \rho(b,\eta,{\bf x}_\perp)/\rho(0,0,0)$, for details see \cite{Ryblewski:2012rr}. The quantity $\varepsilon_{\rm i}$ is the initial energy density at the center of the system created in the most central collisions. Its value is fixed by the measured multiplicity, separately for two different physical scenarios considered in this paper. We use $\varepsilon_{\rm i}$ = 48.8 GeV/fm$^3$ and 80.1 GeV/fm$^3$ for $x_0 = 100$ and $x_0 = 0.032$, respectively.

The evolution is determined by the hydrodynamic equations until the entropy density drops to $\sigma_{\rm f} = 1.79$ fm$^{-3}$, which for $x=1$ corresponds to the temperature $T_{\rm f} = 150$ MeV. According to the single-freeze-out scenario, at this moment the abundances and momenta of particles are expected to be fixed. The processes of particle production and decays of unstable resonances are described by using  {\tt THERMINATOR} \cite{Kisiel:2005hn,Chojnacki:2011hb}, which applies the Cooper-Frye formalism to generate hadrons on the freeze-out hypersurface.

\begin{figure}[t]
\includegraphics[angle=0,width=0.5\textwidth]{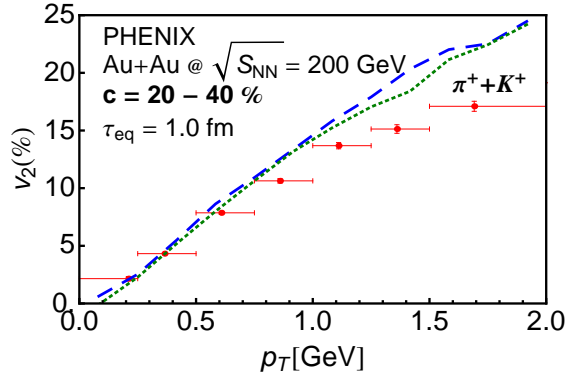}
\caption{\small (Color online) Transverse-momentum dependence of the elliptic flow coefficient $v_2$ of $\pi^{+}+K^{+}$ calculated for $c=20-40$\% ($b=7.84$ fm) at midrapidity and for $\tau_{\rm eq}= 1.0$ fm; $x_0=100$ (dashed blue lines) and $x_0=0.032$ (dotted green lines). The results are compared to the PHENIX Collaboration data (red dots) \cite{Adler:2003kt}. The presented errors are statistical. Horizontal bars denote the $p_T$ bins.
}
\label{fig:ellflowpT}
\end{figure}
\begin{figure}[t]
\includegraphics[angle=0,width=0.5\textwidth]{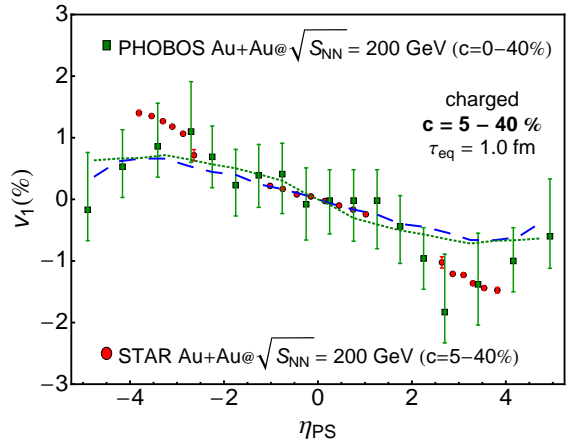}
\caption{\small (Color online) Pseudorapidity dependence of the directed flow of charged particles for the centrality bin $c=5-40$\% and $\tau_{\rm eq}=1.0$ fm;  $x_0=100$ (dashed blue lines) and $x_0=0.032$ (dotted green lines). The results are compared to the experimental data from STAR (red dots) \cite{Abelev:2008jga} and PHOBOS (green squares) \cite{Back:2005pc}. 
}
\label{fig:dirfloweta12}
\end{figure}

\begin{figure}[t]
\includegraphics[angle=0,width=0.5\textwidth]{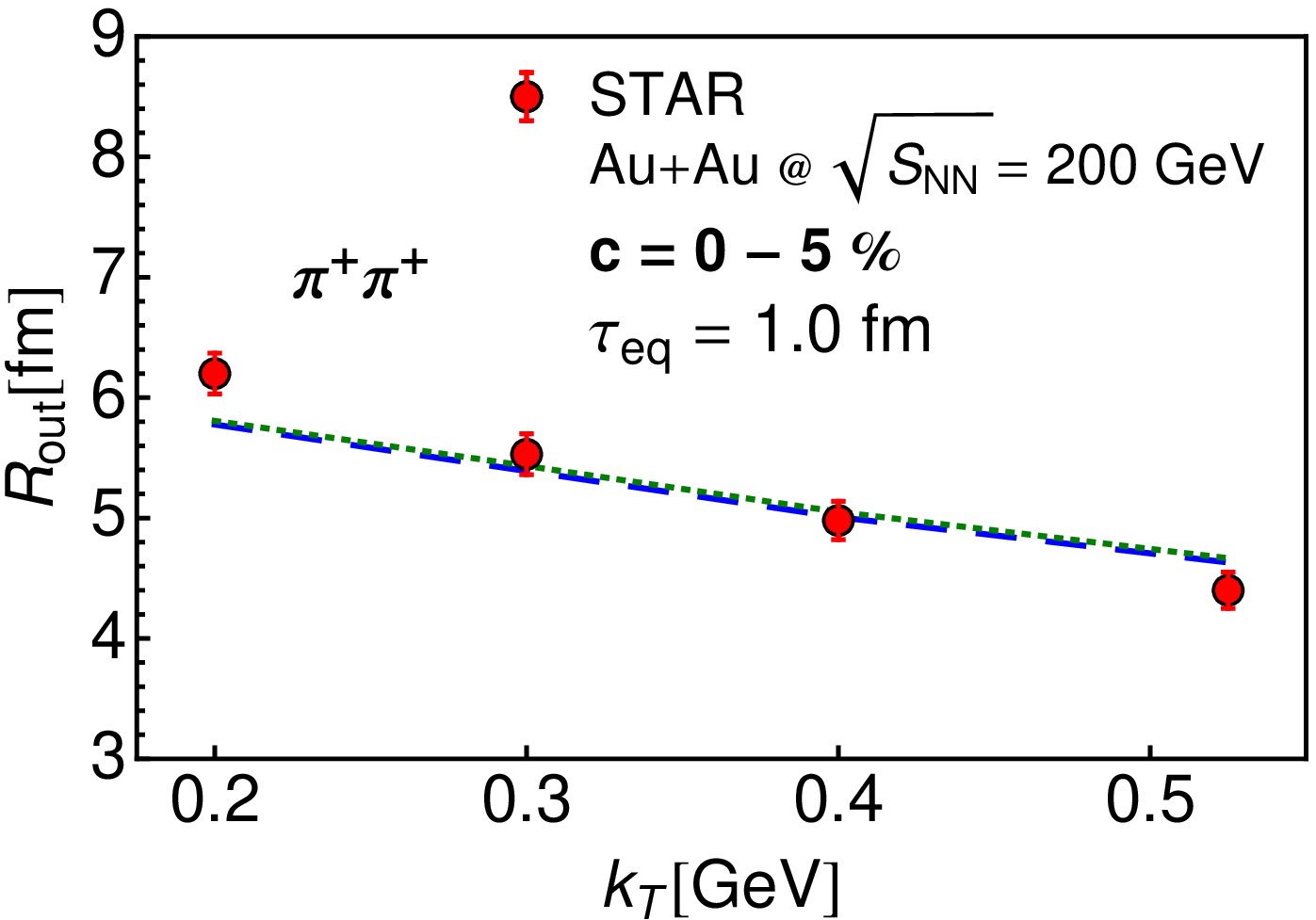}
\caption{\small (Color online) The HBT correlation radius $R_{\rm out}$ of positive pions as a function of the total transverse momentum of the pair for the centrality $c=0-5$\%, $\tau_{\rm eq}=1.0$ fm, and for two values of the initial anisotropy: $x_0=100$ (dashed blue lines) and $x_0=0.032$ (dotted green lines). The results are compared to experimental data from STAR Collaboration (red dots) \cite{Adams:2004yc}. 
}
\label{fig:hbt11}
\end{figure}

\section{Results}

Our results are shown and compared to the RHIC data (Au+Au collisions at the highest beam energy $\sqrt{s_{\rm NN}}$ = 200 GeV) in Figs.~\ref{fig:etadistr_RHIC}--\ref{fig:hbt11}. The model results describing the psudorapidity distributions, transverse-momentum spectra, the elliptic and directed flow coefficients, and the HBT radii have been obtained with different initial anisotropies of pressure; $x_0=100$ (dashed blue lines) and $x_0=0.032$ (dotted green lines). In the all considered cases we find a good agreement between the model results and the data, for more results see \cite{Ryblewski:2012it}. Moreover, we find that the results obtained with different initial anisotropies are practically the same. This is so because we have adjusted the initial energy density separately for two different values of $x_0$. A larger (smaller) initial energy density is used for the initially prolate (oblate) system. 

Our results indicate that the final hadronic observables are not sensitive to the early anisotropy of pressure. The flows are being built during the whole time evolution of the system, hence the relatively short early anisotropic stage does not influence the results. In our opinion, the insensitivity of the hadronic observables helps us to circumvent the early thermalization puzzle.


\begin{theacknowledgments}
WF thanks the organizors of the XII Hadron Physics workshop (April 22--27, 2012, Bento Gonçalves, Brazil) for very kind hospitality. This work was supported by the Polish Ministry of Science and Higher Education under Grant No.~N N202 263438.  
\end{theacknowledgments}



\bibliographystyle{aipproc}   

\end{document}